\documentclass[aps,pre,reprint,longbibliography,floatfix,superscriptaddress]{revtex4-2}
\usepackage{graphicx}
\usepackage{amsmath,amssymb,amsfonts}
\usepackage{hyperref}

\begin{document}
\title{Characterization of domain formation during random sequential adsorption of stiff linear $k$-mers onto a square lattice}

\author{Mikhail V. Ulyanov}
\affiliation{V. A. Trapeznikov Institute of Control Sciences of RAS, Moscow 117997, Russia}
\affiliation{Computational Mathematics and Cybernetics, M. V. Lomonosov Moscow State University, Moscow 119991, Russia}

\author{Yuri Yu. Tarasevich}
\email[Corresponding author: ]{tarasevich@asu.edu.ru}

\author{Andrei V. Eserkepov}
\affiliation{Laboratory of Mathematical Modeling, Astrakhan State University, Astrakhan 414056, Russia}

\author{Irina V. Grigorieva}
\affiliation{Kemerovo State University, Kemerovo 650000, Russia}

\date{\today}

\begin{abstract}
Using computer simulation, we have studied the random sequential adsorption of stiff linear $k$-mers onto a square lattice. Each such particle occupies $k$ adjacent lattice sites. During deposition, the two mutually perpendicular orientations of the particles are equiprobable, hence, a macroscopically isotropic monolayer is formed. However, this monolayer is locally anisotropic, since the deposited particles tend to form domains of particles with the same orientation. Using the ``excluded area'' concept, we have classified lattice sites into several types and examined how the fraction of each type of lattice site varies as the number of deposited particles increases.  The behaviors of these quantities have allowed us to identify the following stages of domain formation (i) the emergence of domain seeds; (ii) the filling of domains; (iii) densification of the domains.
\end{abstract}

\maketitle

\section{Introduction\label{sec:intro}}
Deposition of large particles (proteins, viruses, bacteria, colloids, and macromolecules) at various interfaces is widespread in both nature and industry~\cite{Evans1993RMP,Dabrowski2001ACIS}.
For instance, proteins usually form monolayers on substrates, since proteins do not adhere to each other~\cite{Feder1980JTB,Adamczyk2011ACIS}. Protein adsorption at solid--liquid interfaces is important in thrombosis, plaque formation, artificial organ failure, and fouling of contact lenses~\cite{Adamczyk2011ACIS,Adamczyk2012COCIS}. Efficient separation and purification of proteins by chromatography, filtration, for biosensing, bioreactors, immunological assays, etc. require  controlled protein deposition~\cite{Adamczyk2011ACIS,Adamczyk2012COCIS,Ciesla2020}.
Adsorption of colloid and bioparticles is important for filtration, electroflotation, separation of toner and ink particles, papermaking, xerography, production of magnetic tapes, etc.~\cite{Adamczyk2003}. In general, adsorbed particles, e.g., biological molecules or polymers, have a nonspherical shape. For example, adsorption of fibrinogen has been studied considering its molecule as a linear chain of touching beads of various sizes~\cite{Adamczyk2011}. Another important field is nanotechnology where elongated nanoparticles (e.g., gold nanorods~\cite{Ahmed2013}, colloidal CdSe/CdS
nanorods~\cite{Persano2010}, silver nanorods~\cite{Aslan2005}) are deposited onto a substrate.

Random sequential adsorption (RSA) is a process during which particles are randomly and irreversibly deposited onto a substrate without overlapping with previously adsorbed particles~\cite{Feder1980JTB}. RSA is a useful model for many physical, chemical, and biological processes~\cite{Evans1993RMP,Talbot2000,Adamczyk2012COCIS,Ciesla2020}. Both continuous and discrete substrates can be considered. A widely used kind of discrete substrate is the square lattice. One of the simplest particle shapes is the so-called $k$-mer (rod, stick, needle, stiff linear chain), i.e., a linear ``molecule'' occupying $k$ adjacent lattice sites. The prohibition of overlapping means a hard-core (excluded volume) interaction between the particles. As particles deposit, first there occurs a percolation phase transition, i.e. the emergence of a cluster that penetrates the whole system. Then, the system reaches a jamming state when any additional deposition of particles is impossible due to the absence of any appropriate empty space to place even one extra particle. Although there are some empty spaces, these holes have inappropriate shapes or sizes to accept a further particle. During the RSA of $k$-mers onto a square lattice, the excluded area effect~\cite{Onsager1949ANYAS} leads to the formation of domains filled with particles all of the same orientation. Domain structures have been observed both at percolation~\cite{Leroyer1994PRB,Kondrat2001PRE,Adamczyk2008JChemPhys} and at jamming~\cite{Manna1991JPhA,BudinskiPetkovic1997PhysA,Vandewalle2000EPJE,Kondrat2001PRE}.

A qualitative description of the domain formation is as follows~\cite{BudinskiPetkovic1997PhysA}. At early stages of adsorption, previously deposited $k$-mers affect insignificantly deposition of newcomers since the system is fairly sparse. Almost each new $k$-mer can be adsorbed in arbitrary orientations. As the number of deposited $k$-mers increases, a newly deposited $k$-mers have to align to the already deposited ones to avoid intersections. The late-stage deposition pattern consists of domains of densely packed parallel $k$-mers and regions of empty sites of sizes ranging from single site to the length $k-1$ which are inaccessible for adsorption of $k$-mers. Similar behavior, i.e.,  formation of domains of parallelly deposited objects, has also been observed for elongated particles of the more complex shapes on both square~\cite{BudinskiPetkovic1997PhysA} and triangular lattices~\cite{BudinskiPetkovic1997PRE}. The sizes of these domains
were greater for more elongated shapes, i.e., for the shapes that resemble more the straight lines. Moreover, domains have been reported for RSA of binary mixtures of line segments on a square lattice~\cite{Lee2000}.

Using a local order parameter, the typical size of domains has been evaluated as $k \times k$~\cite{Tarasevich2019JPCS}. Thus, although a monolayer produced by RSA is macroscopically isotropic, microscopic regions can exhibit significant anisotropy. Visually, the domains look like winding areas with dense centers and diffuse edges~\cite{Slutskii2018PRE}. It seems, therefore, that a local order parameter cannot provide complete information about the shape and structure of the domains. An alternative characteristic of the domain structure is the pair correlation function~\cite{Ricci1994JChemPhys,Hart2016PRE}.

Internal structure of domains can be described using a concept of stacks~\cite{Kundu2013PRE}.
A definition of stack states that two neighboring parallel $k$-mers belong to the same stack if the number of nearest-neighbor bonds between them is greater than $k/2$~\cite{Kundu2013PRE}. When a stack is defined in such a way, it has a wormlike structure without branching. Although all $k$-mers within the stack are aligned in the same direction, some transverse fluctuations are allowed, which make stacks wavy and curved. Accordingly, each domain is a set of stacks.

A larger substructure of a domain is a cluster of $k$-mers of the same orientation. Two neighboring parallel $k$-mers belong to the same cluster if there is at least one  nearest-neighbor bond between them. This substructure has been used to characterize a relaxation of the jammed state due to random walks of $k$-mers~\cite{Tarasevich2017JSM}. Thus, the contiguous stacks form a cluster.

When RSA of randomly oriented elongated particles onto a continuous substrate is considered, formation of domains has been observed in the long-time regime for the zero-width sticks~\cite{Sherwood1990JPhA}, rectangles~\cite{Vigil1989JChemPhys}, discorectangles~\cite{Ricci1994JChemPhys}, and polymers~\cite{Ciesla2013PRE}. These elongated particles were arranged almost parallel to each other within domains. Thus, formation of domains is custom for elongated particles when deposited onto both discrete and continuous substrates.

Recently, a RSA of rectangles onto a continuous plane has been considered geometrically~\cite{Tsiantis2019PhysA}. The three areas which are formed around a deposited rectangle have been defined. As particles are deposited they change the properties of the surrounding space by creating a probability field around them, i.e., create a ``polarized space''. Within this ``polarized space'', a newly deposited particle is forced to align parallel to the previously deposited particles. The proposed approach can be treated as a refinement of the excluded area concept. Similar consideration has been performed for zero-width sticks, i.e., rectangles with infinity large aspect ratio~\cite{Heitz2011NanoTech}.

By means of both computer simulation and analytical treatment using the ``excluded area'' concept, we have studied the formation of domains during RSA of $k$-mers onto a square lattice. The three  stages have been found and classified. In fact, we have transferred the idea of a ``polarized space''~\cite{Tsiantis2019PhysA} from the continuous space to the discrete one.

The rest of the paper is constructed as follows. In Sec.~\ref{sec:methods}, the technical details of the simulations are described, all necessary quantities are defined, and some estimates of the finite-size effect are given. Section~\ref{sec:results} presents our principal findings. Section~\ref{sec:conclusion} summarizes the main results.

\section{Computational model\label{sec:methods}}
A square lattice with $L \times L$ sites was used as a substrate. Periodic boundary conditions were  applied along both directions of the lattice to reduce the finite-size effect. Linear particles occupying $k$ adjacent lattice sites were randomly and sequentially deposited onto the lattice. To distinguish the two possible orientations of deposited particles, we denoted the particles oriented along the abscissa as $k_x$-mers, while the particles oriented along the ordinate were $k_y$-mers.  We treated the leftmost site of a $k_x$-mer and the topmost site of a $k_y$-mer as the ``primary element'' (origin) of the particle. The rest of the $k-1$ sites of the particle were denoted as its body. Both the mutually perpendicular orientations of deposited particles taken as equiprobable. In our simulations, we used $k \in [2;12]$. As a basis, the linear size of the lattice was chosen as $L = 32 k$. However, the finite-size effect has also been tested by variations of the lattice size for a fixed value of $k$. All results were averaged over 100 independent runs.

We used the reduced (normalized) coverage, i.e., the number of occupied sites, $N$, divided by the number of occupied sites at jamming, $N_\text{j}$,
\begin{equation}\label{eq:x}
  x = \frac{N}{N_\text{j}},
\end{equation}
in such a way that $x \in [0;1]$.

\begin{figure}[!htb]
  \centering
  \includegraphics[width=\columnwidth]{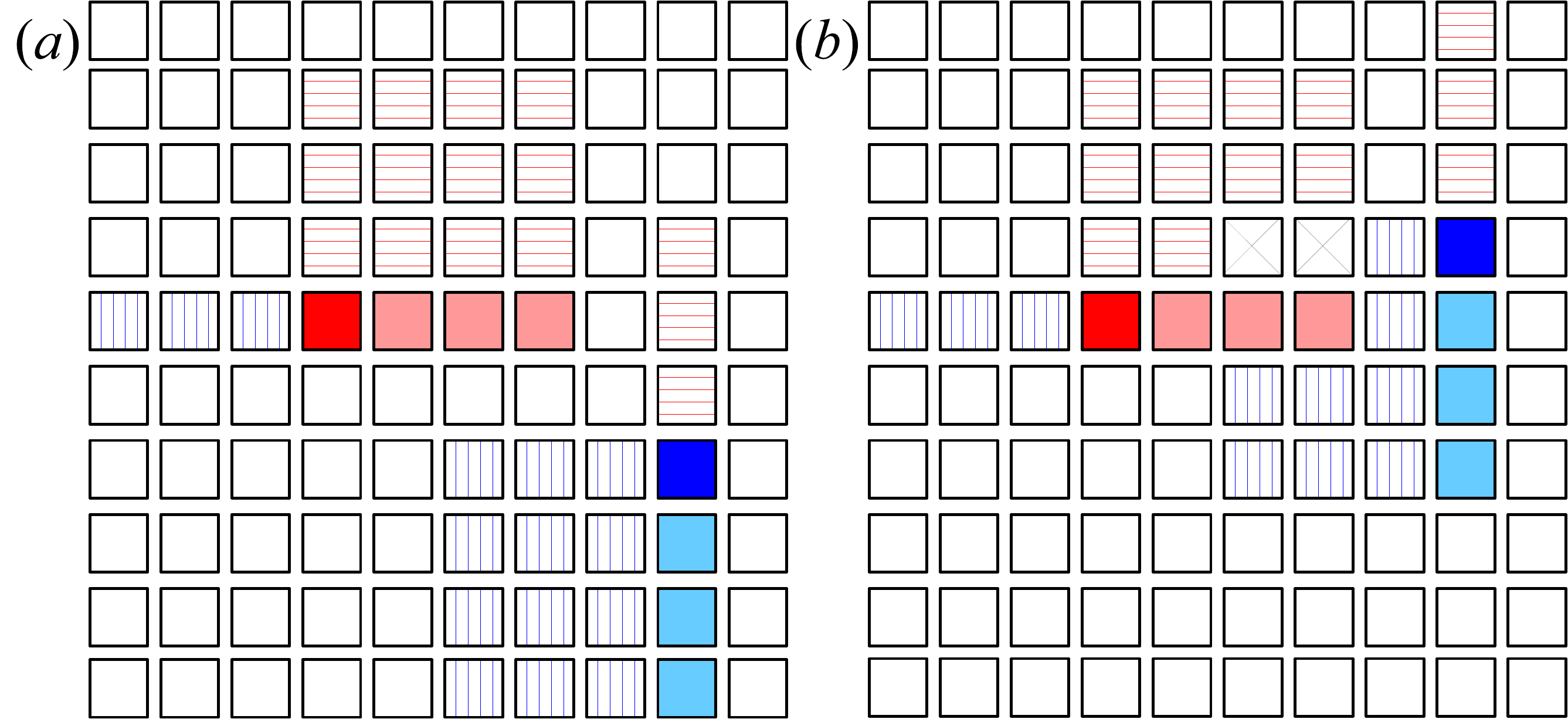}
  \caption{Example of a $k_x$-mer and a $k_y$-mer ($k=4$) with (a) non-overlapping excluded areas and (b) partially overlapping excluded areas.\label{fig:lattice}}
\end{figure}

Each adsorbed particle blocks $k$ lattice sites from further deposition of both $k_x$- and $k_y$-mers. Furthermore, some sites in the vicinity of the adsorbed particle are forbidden for the deposition of only one kind of particle (Fig.~\ref{fig:lattice}). Figure~\ref{fig:lattice}(a) demonstrates a $k_x$-mer and a $k_y$-mer together with their non-overlapping excluded areas. Figure~\ref{fig:lattice}(b) demonstrates a $k_x$-mer and a $k_y$-mer when their excluded areas are partially overlapping. Deposited particles are shown using solid fill. Darker cells correspond to the ``primary element'' of particles, while lighter ones form their bodies. The ``primary elements'' of any additional $k_x$- or $k_y$-mers can be placed in open cells. Only $k_y$-mer ``primary elements'' can be placed in cells with vertical hatching. The ``primary elements'' of only $k_x$-mers can be placed in cells with horizontal hatching. The ``primary elements'' of neither $k_x$- nor $k_y$-mers can be placed into cross-hatched cells.

We classified each of the lattice sites under one of several types:
\begin{description}
\item[Type 0] Lattice sites that are forbidden for the deposition of both $k_x$- and $k_y$-mers. This type can be additionally divided into two subtypes:
    \begin{description}
      \item[Subtype $-0$] Occupied sites [filled squares in Fig.~\ref{fig:lattice}(a)]. No site of the newly deposited particle can be placed in these sites.
      \item[Subtype $+0$] Empty sites that are forbidden for the deposition of the ``primary elements'' of both $k_x$- and $k_y$-mers. However, a body-site of a newly deposited particle may be placed into a site of subtype $+0$. These sites are shown in Fig.~\ref{fig:lattice}(b) as cross-hatched squares.
    \end{description}
\item[Type 1] Empty sites that allow deposition of the ``primary elements'' of the either $k_x$- or $k_y$-mers. These sites are shown in Fig.~\ref{fig:lattice} as horizontally, or vertically hatched squares, respectively.
\item[Type 2] Empty sites that can allow the deposition of the ``primary elements'' of both $k_x$- and $k_y$-mers. These sites are shown in Fig.~\ref{fig:lattice} as open squares.
\end{description}
Sites of types 0 and 1 belong to the excluded area.

The fractions of lattice sites belonging to one of these types are denoted as $f_{-0}(x)$, $f_{+0}(x)$, $f_1(x)$, and $f_2(x)$, respectively. Naturally, $f_{-0}(x) + f_{+0}(x) + f_1(x) + f_2(x) = 1$, hence, only three of the four functions are independent. By definition, $f_{-0}(x)$  is a linear function.  It is therefore uninformative, and is not discussed further.

\begin{figure}[!htb]
  \centering
  \includegraphics[width=\columnwidth]{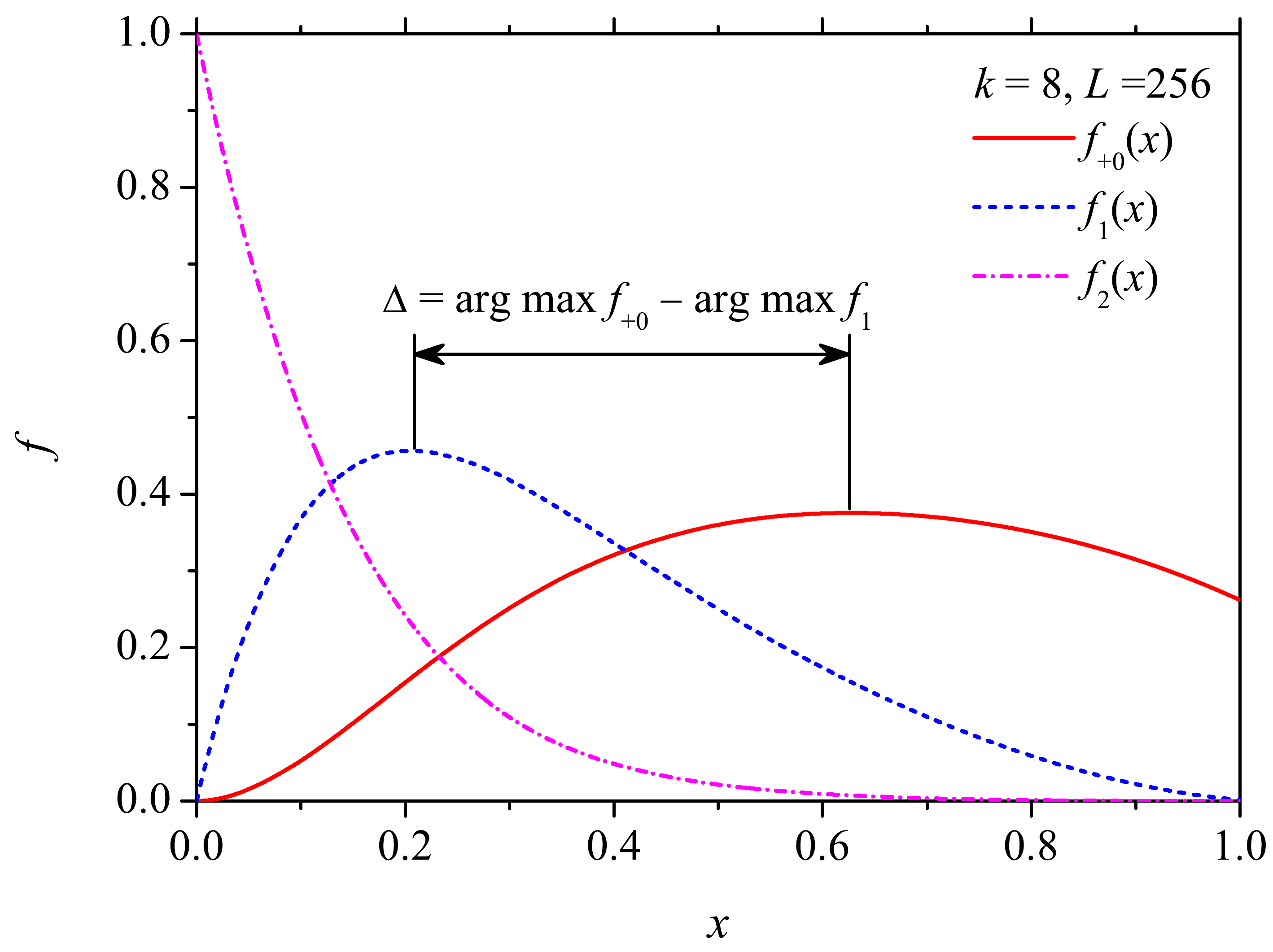}
  \caption{Example of the functions $f_{+0}(x), f_1(x)$, and $f_2(x)$ for $k=8$, $L=256$.\label{fig:fk8L256}}
\end{figure}

Figure~\ref{fig:fk8L256} presents an example of the functions $f_{+0}(x)$, $f_1(x)$, and $f_2(x)$ for one particular case ($k=8$, $L=256$). $f_2(x)$ is a monotonically decreasing function, while each of the functions $f_{+0}(x)$ and $f_1(x)$ has one maximum and one inflection point. The coordinates of the maxima look promising for characterizing the kinetics of domain formation.

\begin{figure}[!htb]
  \centering
  \includegraphics[width=\columnwidth]{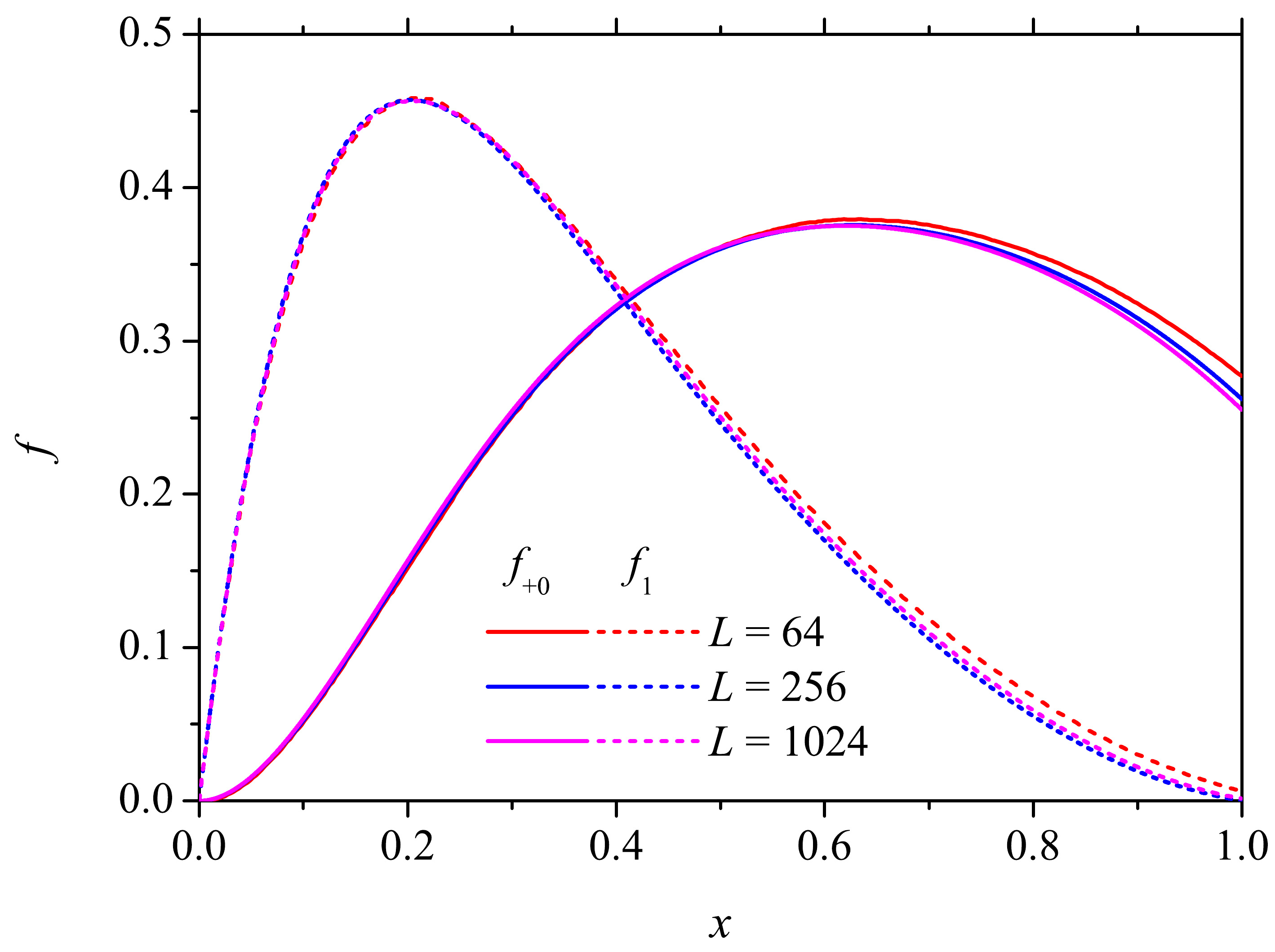}
  \caption{Example of the finite-size effect: functions $f_{+0}(x)$ and $f_1(x)$ for $k=8$, $L=64, 256, 1024$.\label{fig:fse}}
\end{figure}

Figure~\ref{fig:fse} presents the functions $f_{+0}(x)$ and $f_1(x)$ for a fixed $k=8$ and different lattice sizes ($L=64, 256, 1024$). Figure~\ref{fig:fse} suggests that the finite-size effect is significant only in the vicinity of the jammed state ($x=1$). In any case, the curves for $L=32k$ and $L=128k$ are hardly distinguishable. Since domain formation is a continuous process, i.e., there is no jump between any two stages, the exact location of the maxima is not so important. This is the reason for the use of $L=32k$ in our main evaluations.

\begin{figure}[!htb]
  \centering
  \includegraphics[width=\columnwidth]{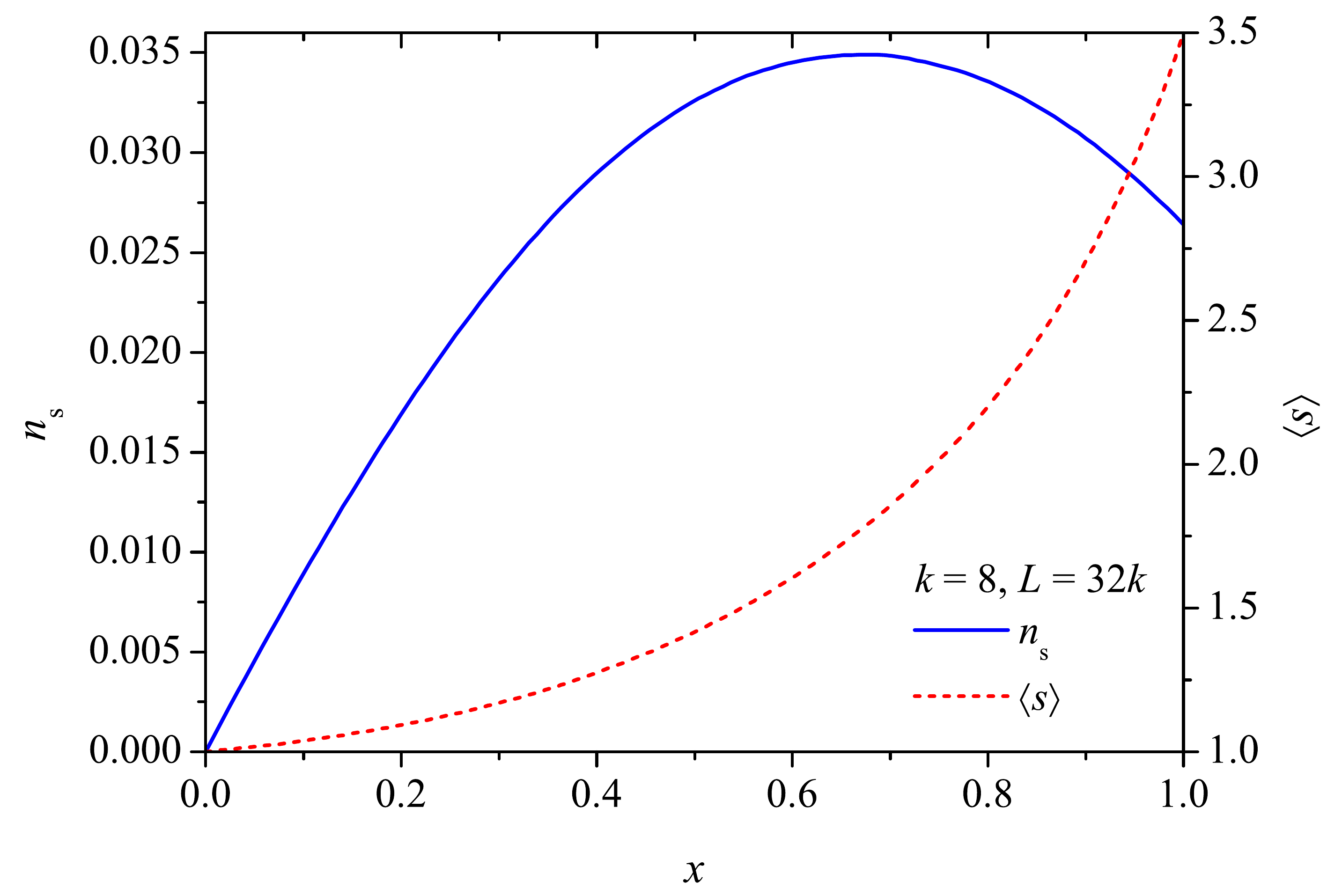}
  \caption{Example of the dependencies of the mean stack size, $\langle s \rangle$, and the number of stacks per lattice site, $n_\text{s}$, on the normalized coverage, $x$, for $k=8$, $L=256$.\label{fig:stackk8}}
\end{figure}

We investigated the mean stack size, $\langle s \rangle$, and the number of stacks per lattice site, $n_\text{s}$, which quantify internal structure of domains. Figure~\ref{fig:stackk8} demonstrates an example of the dependencies of these quantities on the normalized coverage, $x$, for $k=8$, $L=256$. Quite expectedly, the mean stack size monotonically increases. The number of stacks per lattice site has a maximum at a certain value of the normalized coverage, $x = \arg\max n_\text{s}(x)$.

\begin{figure}[!htb]
  \centering
  \includegraphics[width=\columnwidth]{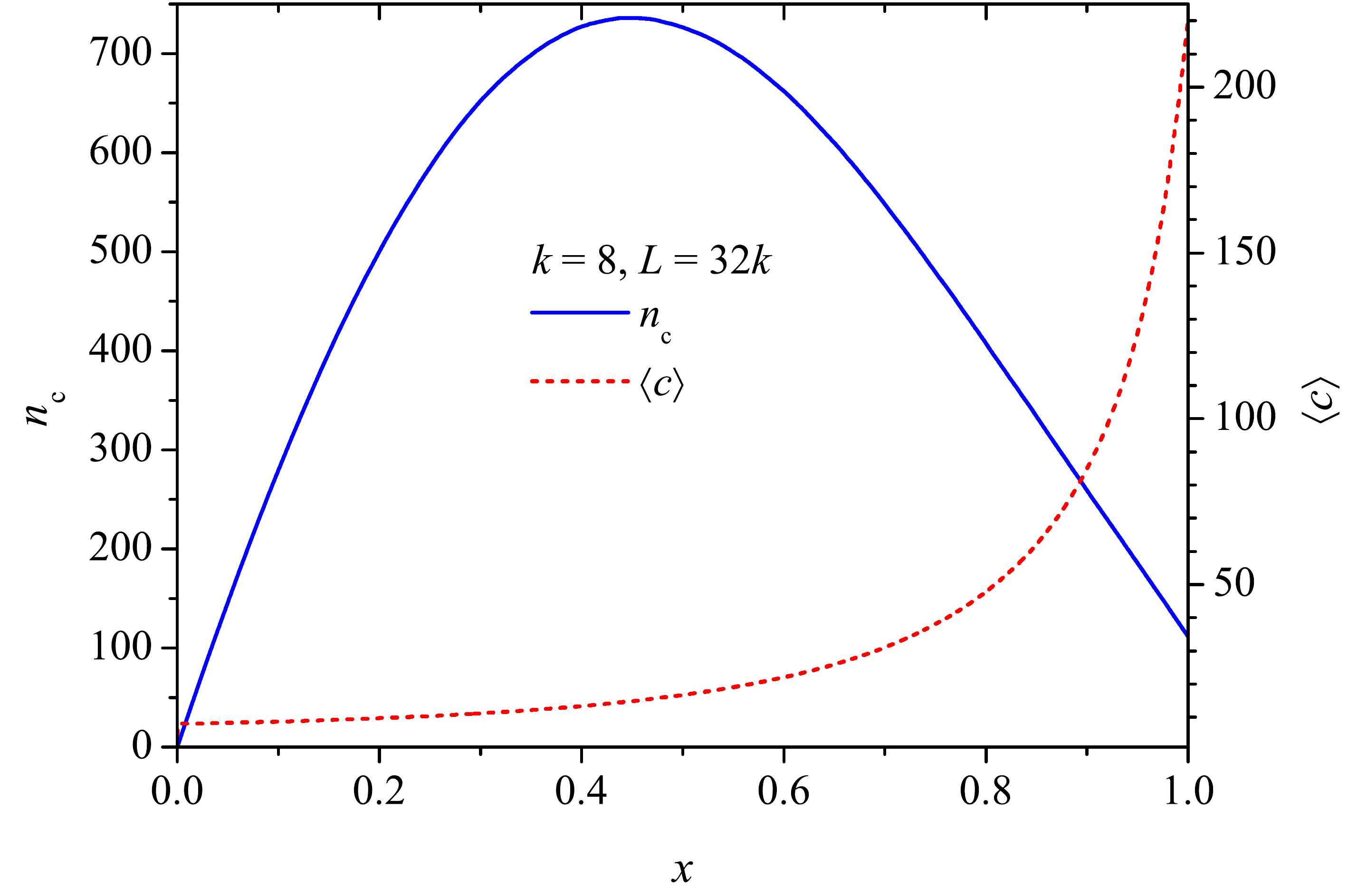}
  \caption{Example of the dependencies of the mean cluster size, $\langle c \rangle$, and the number of clusters per lattice site, $n_\text{c}$, on the normalized coverage, $x$, for $k=8$, $L=256$.\label{fig:clusterk8}}
\end{figure}

Additionally, we investigated the mean cluster size, $\langle c \rangle$, and the number of clusters per lattice site, $n_\text{s}$, which quantify internal structure of domains. Figure~\ref{fig:clusterk8} demonstrates an example of the dependencies of these quantities on the normalized coverage, $x$, for $k=8$, $L=256$. As expected, the mean cluster size monotonically increases. The number of clusters per lattice site has a maximum at a certain value of the normalized coverage, $x = \arg\max n_\text{c}(x)$.

\section{Results and Discussion\label{sec:results}}
\begin{figure}[!htb]
  \centering
  \includegraphics[width=\columnwidth]{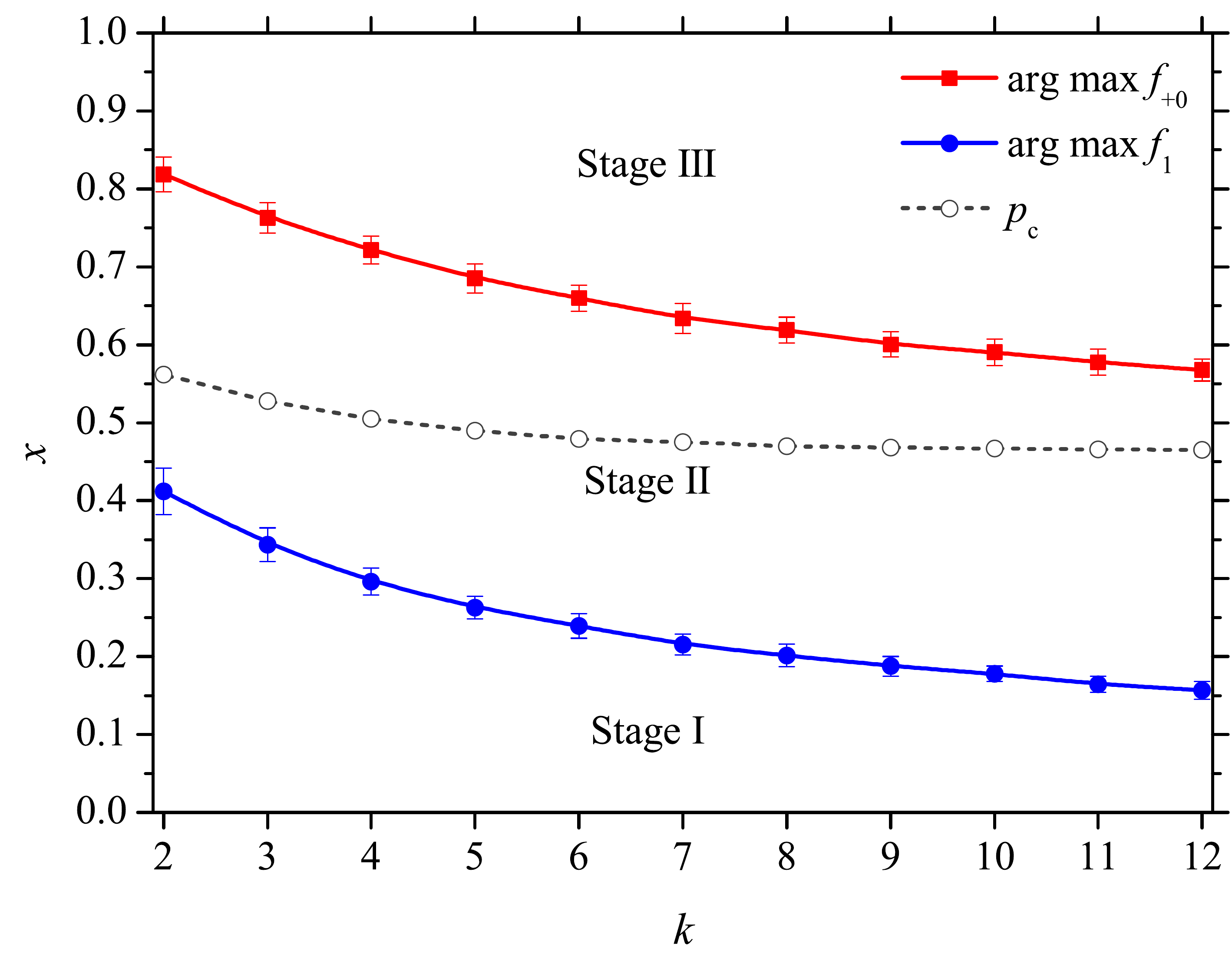}
  \caption{Dependencies of $\arg\max f_{+0}$, $\arg\max f_1$, $\arg\max n_\text{s}$, and $\arg\max n_\text{c}$ on the particle size, $k$. The lines between the markers are drawn simply for convenience. The error bars correspond to the standard deviation. When not
shown explicitly, they are of the order of the marker size.\label{fig:criticalpoints}}
\end{figure}

For  $k \in [2;12]$, the abscissae of the extremal points of the functions $f_1(x)$ and $f_{+0}(x)$ decrease as the value of $k$ increases (Fig.~\ref{fig:criticalpoints}). However, this behavior may differ for larger values of $k$.

\begin{figure}[!htb]
  \centering
  \includegraphics[width=0.65\columnwidth]{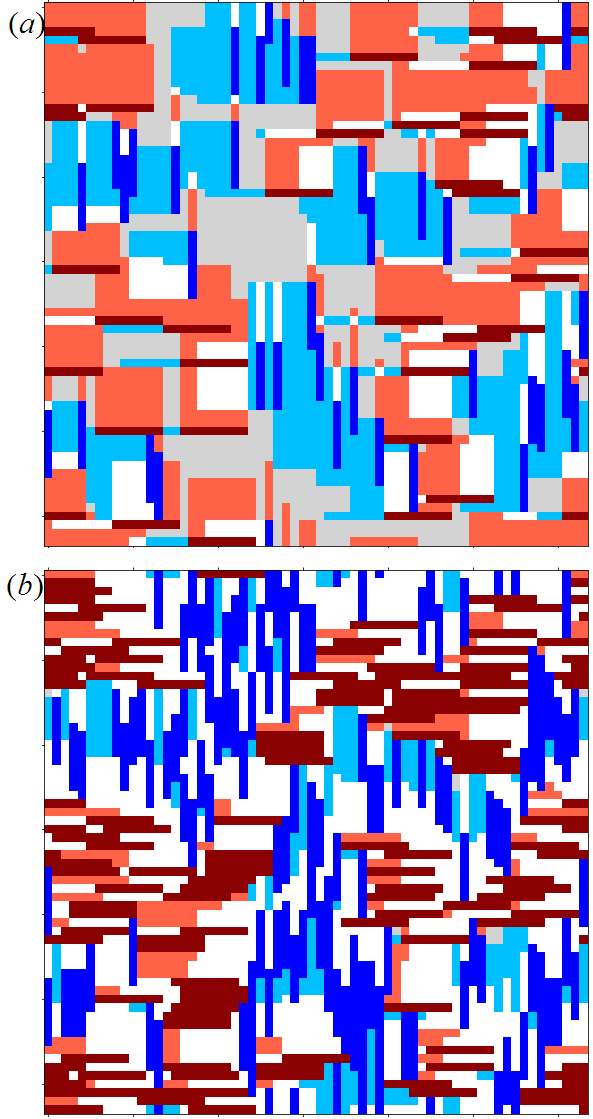}
  \caption{Example of a system under consideration (a) at the end of stage I and (b) at the end of stage II; $k=8$, $L=8k$. Both deposited particles and different types of empty sites are shown. Darker regions correspond to deposited particles (sites of subtype $-0$). Empty sites belonging to subtype $+0$ are shown as white regions. Light gray regions correspond to type 2 sites. Two shades of gray (light blue and light red online) depict sites belonging to type 1.\label{fig:stage1-2}}
\end{figure}

The extremal points of the functions $f_1(x)$ and $f_{+0}(x)$ as well as direct observation of the particle deposition~\footnote{See Supplemental Material at [URL will be inserted by publisher] for an animation of the temporal evolution of domains for $k=8$, $L=16k$.} suggest the following stages of domain formation. Naturally, the boundaries of the stages are approximate (Fig.~\ref{fig:criticalpoints}).
\paragraph*{Stage I: Emergence of domain seeds.} During the initial stage of particle deposition when $x \in [0; \arg \max f_1(x)]$, particles stake out the future domains. The number of empty sites of type 2 decreases, while the number of empty sites of type 1 increases. Since a significant fraction of the empty sites can accept deposited particles of only one orientation, they can be treated as the progenitors of future domains [Fig.~\ref{fig:stage1-2}(a)].

\paragraph*{Stage II: Filling of domains.} As the number of deposited particles increases, $x \in [\arg \max f_1(x); \arg \max f_{+0}(x)]$, the number of sites of type 1 decreases due to overlapping of the excluded areas produced by the deposited particles of mutually perpendicular orientations [Fig.~\ref{fig:stage1-2}(b)].

\begin{figure*}[!tb]
  \centering
  \includegraphics[width=\textwidth]{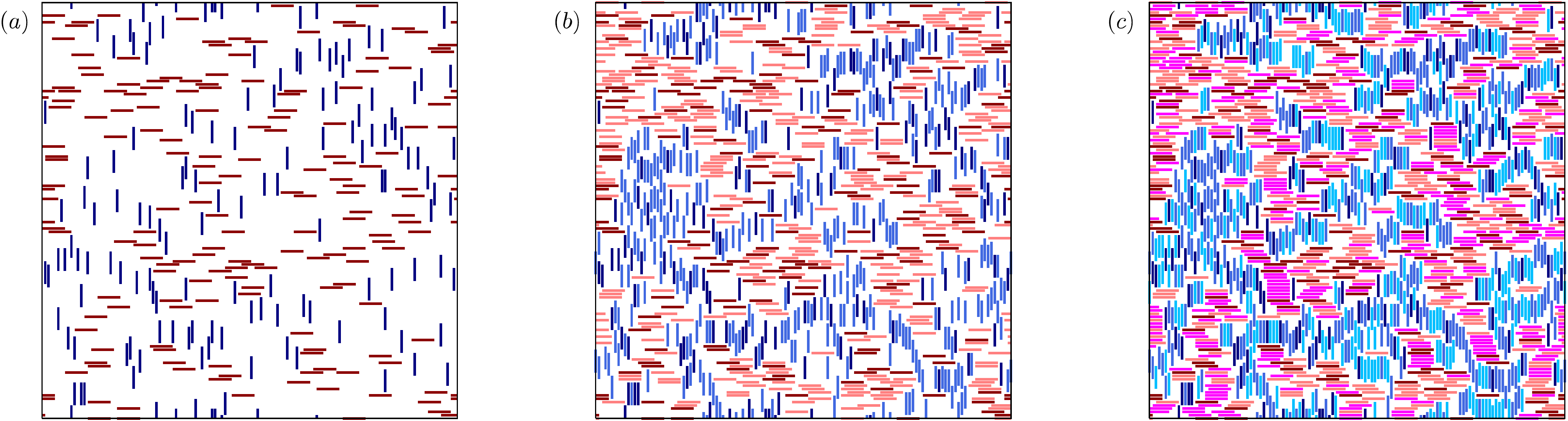}
  \caption{Example of the sequence of formation of the domain structure for one particular lattice ($k=8$, $L=16k$). The patterns at the end of each stage are shown. (a) I, (b) II, and (c) III. Particles deposited during each particular stage are shown in different shades. \label{fig:stages}}
\end{figure*}

\paragraph*{Stage III: Densification of domains.} At this stage when $x \in [\arg \max f_{+0}(x); 1]$, almost all newly deposited particles fall only into the already formed domains. A feature of this stage is the reduction in the number of sites of type $+0$ due to their overlapping by newly deposited particles. At this stage, the number of sites of type 2 is already negligible. A  reduction in the number of sites of type 1 occurs since the newly deposited particles overlap sites of types 1 and +0. Almost all newly deposited particles are placed into already formed and limited domain structures. This densification of the domains little changes their formed structure, since almost all the newly deposited particles are placed inside domains between, and aligned with, previously placed particles [Fig.~\ref{fig:stages}(c)].

For the values of $k$ under consideration, the width of stage II seems to be a constant within the precision of our evaluations
$$
\Delta = \arg \max f_{+0} - \arg \max f_{1} \approx 0.42 \pm 0.01.
$$
With increasing value of $k$, the width of stage I decreases while that of stage III increases  (Fig.~\ref{fig:criticalpoints}).

During Stage II, the number of clusters reaches its maximum value and then decreases (Fig.~\ref{fig:criticalpoints}). Merging of clusters seems to be independent of evolution of the ``polarized space''.

The number of stacks reaches its maximum value near the transition from Stage II to Stage II and then decreases (Fig.~\ref{fig:criticalpoints}). Non-monotonic dependence $n_\text{s}(k)$ occurs due to definition of stacks~\cite{Kundu2013PRE} since there are the two different branches corresponding to even and odd values of $k$. For example, according to the definition of stacks~\cite{Kundu2013PRE}, two common lateral bonds are needed for two particles to belong to the same stack both in the case of dimers and trimers.

\section{Conclusion\label{sec:conclusion}}

Using computer simulation, we have studied an isotropic random sequential adsorption of stiff linear segments ($k$-mers) onto a square lattice with periodic boundary conditions along both  directions. Due to the excluded area effect, deposited particles form domains of particles of the same orientation. Using the excluded area concept, we have classified  lattice sites into several types. We have examined how the fraction of each type of lattice site varies with the number of deposited particles. The behaviors of these quantities provide for a classification of the stages of domain formation: (i) the emergence of domain seeds [Fig.~\ref{fig:stages}(a)]; (ii) the filling of domains [Fig.~\ref{fig:stages}(b)]; (iii) densification of the domains [Fig.~\ref{fig:stages}(c)]. Our approach and results are closely related to that for RSA of needles~\cite{Heitz2011NanoTech} and rectangles~\cite{Tsiantis2019PhysA} onto a plane. Since our computer simulation is restricted only to short particles, an additional study is needed for larger values of $k$ ($k>12$); however, such study is expected to be time-consuming.

Our study offers an approach to classify the RSA stages.
The proposed approach is expected to be useful for other kinds of regular discrete substrates (e.g., triangular lattice) and other shapes of particles (e.g., rectangles). Application of the approach to other kinds of substrates as well as to other particle shapes suggests an additional independent study.

\acknowledgments
We acknowledge funding from the Russian Foundation for Basic Research, Project No.~18-07-00343.

\bibliography{RSA}

\begin{thebibliography}{33}%
\makeatletter
\providecommand \@ifxundefined [1]{%
 \@ifx{#1\undefined}
}%
\providecommand \@ifnum [1]{%
 \ifnum #1\expandafter \@firstoftwo
 \else \expandafter \@secondoftwo
 \fi
}%
\providecommand \@ifx [1]{%
 \ifx #1\expandafter \@firstoftwo
 \else \expandafter \@secondoftwo
 \fi
}%
\providecommand \natexlab [1]{#1}%
\providecommand \enquote  [1]{``#1''}%
\providecommand \bibnamefont  [1]{#1}%
\providecommand \bibfnamefont [1]{#1}%
\providecommand \citenamefont [1]{#1}%
\providecommand \href@noop [0]{\@secondoftwo}%
\providecommand \href [0]{\begingroup \@sanitize@url \@href}%
\providecommand \@href[1]{\@@startlink{#1}\@@href}%
\providecommand \@@href[1]{\endgroup#1\@@endlink}%
\providecommand \@sanitize@url [0]{\catcode `\\12\catcode `\$12\catcode
  `\&12\catcode `\#12\catcode `\^12\catcode `\_12\catcode `\%12\relax}%
\providecommand \@@startlink[1]{}%
\providecommand \@@endlink[0]{}%
\providecommand \url  [0]{\begingroup\@sanitize@url \@url }%
\providecommand \@url [1]{\endgroup\@href {#1}{\urlprefix }}%
\providecommand \urlprefix  [0]{URL }%
\providecommand \Eprint [0]{\href }%
\providecommand \doibase [0]{https://doi.org/}%
\providecommand \selectlanguage [0]{\@gobble}%
\providecommand \bibinfo  [0]{\@secondoftwo}%
\providecommand \bibfield  [0]{\@secondoftwo}%
\providecommand \translation [1]{[#1]}%
\providecommand \BibitemOpen [0]{}%
\providecommand \bibitemStop [0]{}%
\providecommand \bibitemNoStop [0]{.\EOS\space}%
\providecommand \EOS [0]{\spacefactor3000\relax}%
\providecommand \BibitemShut  [1]{\csname bibitem#1\endcsname}%
\let\auto@bib@innerbib\@empty
\bibitem [{\citenamefont {Evans}(1993)}]{Evans1993RMP}%
  \BibitemOpen
  \bibfield  {author} {\bibinfo {author} {\bibfnamefont {J.~W.}\ \bibnamefont
  {Evans}},\ }\bibfield  {title} {\bibinfo {title} {Random and cooperative
  sequential adsorption},\ }\href {https://doi.org/10.1103/RevModPhys.65.1281}
  {\bibfield  {journal} {\bibinfo  {journal} {Rev. Mod. Phys.}\ }\textbf
  {\bibinfo {volume} {65}},\ \bibinfo {pages} {1281} (\bibinfo {year}
  {1993})}\BibitemShut {NoStop}%
\bibitem [{\citenamefont {D\c{a}browski}(2001)}]{Dabrowski2001ACIS}%
  \BibitemOpen
  \bibfield  {author} {\bibinfo {author} {\bibfnamefont {A.}~\bibnamefont
  {D\c{a}browski}},\ }\bibfield  {title} {\bibinfo {title} {Adsorption --- from
  theory to practice},\ }\href {https://doi.org/10.1016/S0001-8686(00)00082-8}
  {\bibfield  {journal} {\bibinfo  {journal} {Adv. Colloid Interface Sci.}\
  }\textbf {\bibinfo {volume} {93}},\ \bibinfo {pages} {135} (\bibinfo {year}
  {2001})}\BibitemShut {NoStop}%
\bibitem [{\citenamefont {Feder}(1980)}]{Feder1980JTB}%
  \BibitemOpen
  \bibfield  {author} {\bibinfo {author} {\bibfnamefont {J.}~\bibnamefont
  {Feder}},\ }\bibfield  {title} {\bibinfo {title} {Random sequential
  adsorption},\ }\href {https://doi.org/10.1016/0022-5193(80)90358-6}
  {\bibfield  {journal} {\bibinfo  {journal} {J. Theor. Biol.}\ }\textbf
  {\bibinfo {volume} {87}},\ \bibinfo {pages} {237} (\bibinfo {year}
  {1980})}\BibitemShut {NoStop}%
\bibitem [{\citenamefont {Adamczyk}\ \emph
  {et~al.}(2011{\natexlab{a}})\citenamefont {Adamczyk}, \citenamefont
  {Nattich}, \citenamefont {Wasilewska},\ and\ \citenamefont
  {Zaucha}}]{Adamczyk2011ACIS}%
  \BibitemOpen
  \bibfield  {author} {\bibinfo {author} {\bibfnamefont {Z.}~\bibnamefont
  {Adamczyk}}, \bibinfo {author} {\bibfnamefont {M.}~\bibnamefont {Nattich}},
  \bibinfo {author} {\bibfnamefont {M.}~\bibnamefont {Wasilewska}},\ and\
  \bibinfo {author} {\bibfnamefont {M.}~\bibnamefont {Zaucha}},\ }\bibfield
  {title} {\bibinfo {title} {Colloid particle and protein deposition ---
  electrokinetic studies},\ }\href {https://doi.org/10.1016/j.cis.2011.04.002}
  {\bibfield  {journal} {\bibinfo  {journal} {Adv. Colloid Interface Sci.}\
  }\textbf {\bibinfo {volume} {168}},\ \bibinfo {pages} {3} (\bibinfo {year}
  {2011}{\natexlab{a}})},\ \bibinfo {note} {surface forces and thin liquid
  films}\BibitemShut {NoStop}%
\bibitem [{\citenamefont {Adamczyk}(2012)}]{Adamczyk2012COCIS}%
  \BibitemOpen
  \bibfield  {author} {\bibinfo {author} {\bibfnamefont {Z.}~\bibnamefont
  {Adamczyk}},\ }\bibfield  {title} {\bibinfo {title} {Modeling adsorption of
  colloids and proteins},\ }\href {https://doi.org/10.1016/j.cocis.2011.12.002}
  {\bibfield  {journal} {\bibinfo  {journal} {Curr. Opin. Colloid Interface
  Sci.}\ }\textbf {\bibinfo {volume} {17}},\ \bibinfo {pages} {173} (\bibinfo
  {year} {2012})}\BibitemShut {NoStop}%
\bibitem [{\citenamefont {Cie\'{s}la}(2020)}]{Ciesla2020}%
  \BibitemOpen
  \bibfield  {author} {\bibinfo {author} {\bibfnamefont {M.}~\bibnamefont
  {Cie\'{s}la}},\ }\bibfield  {title} {\bibinfo {title} {Effective modelling of
  adsorption monolayers built of complex molecules},\ }\href
  {https://doi.org/10.1016/j.jcp.2019.108999} {\bibfield  {journal} {\bibinfo
  {journal} {J. Comput. Phys.}\ }\textbf {\bibinfo {volume} {401}},\ \bibinfo
  {pages} {108999} (\bibinfo {year} {2020})}\BibitemShut {NoStop}%
\bibitem [{\citenamefont {Adamczyk}(2003)}]{Adamczyk2003}%
  \BibitemOpen
  \bibfield  {author} {\bibinfo {author} {\bibfnamefont {Z.}~\bibnamefont
  {Adamczyk}},\ }\bibfield  {title} {\bibinfo {title} {Particle adsorption and
  deposition: role of electrostatic interactions},\ }\href
  {https://doi.org/10.1016/s0001-8686(02)00062-3} {\bibfield  {journal}
  {\bibinfo  {journal} {Adv. Colloid Interface Sci.}\ }\textbf {\bibinfo
  {volume} {100-102}},\ \bibinfo {pages} {267} (\bibinfo {year}
  {2003})}\BibitemShut {NoStop}%
\bibitem [{\citenamefont {Adamczyk}\ \emph
  {et~al.}(2011{\natexlab{b}})\citenamefont {Adamczyk}, \citenamefont
  {Barbasz},\ and\ \citenamefont {Cie{\'s}la}}]{Adamczyk2011}%
  \BibitemOpen
  \bibfield  {author} {\bibinfo {author} {\bibfnamefont {Z.}~\bibnamefont
  {Adamczyk}}, \bibinfo {author} {\bibfnamefont {J.}~\bibnamefont {Barbasz}},\
  and\ \bibinfo {author} {\bibfnamefont {M.}~\bibnamefont {Cie{\'s}la}},\
  }\bibfield  {title} {\bibinfo {title} {Mechanisms of fibrinogen adsorption at
  solid substrates},\ }\href {https://doi.org/10.1021/la200798d} {\bibfield
  {journal} {\bibinfo  {journal} {Langmuir}\ }\textbf {\bibinfo {volume}
  {27}},\ \bibinfo {pages} {6868} (\bibinfo {year}
  {2011}{\natexlab{b}})}\BibitemShut {NoStop}%
\bibitem [{\citenamefont {Ahmed}\ \emph {et~al.}(2013)\citenamefont {Ahmed},
  \citenamefont {Glass}, \citenamefont {Kooij},\ and\ \citenamefont {van
  Ruitenbeek}}]{Ahmed2013}%
  \BibitemOpen
  \bibfield  {author} {\bibinfo {author} {\bibfnamefont {W.}~\bibnamefont
  {Ahmed}}, \bibinfo {author} {\bibfnamefont {C.}~\bibnamefont {Glass}},
  \bibinfo {author} {\bibfnamefont {E.~S.}\ \bibnamefont {Kooij}},\ and\
  \bibinfo {author} {\bibfnamefont {J.~M.}\ \bibnamefont {van Ruitenbeek}},\
  }\bibfield  {title} {\bibinfo {title} {Tuning the oriented deposition of gold
  nanorods on patterned substrates},\ }\href
  {https://doi.org/10.1088/0957-4484/25/3/035301} {\bibfield  {journal}
  {\bibinfo  {journal} {Nanotechnology}\ }\textbf {\bibinfo {volume} {25}},\
  \bibinfo {pages} {035301} (\bibinfo {year} {2013})}\BibitemShut {NoStop}%
\bibitem [{\citenamefont {Persano}\ \emph {et~al.}(2010)\citenamefont
  {Persano}, \citenamefont {Giorgi}, \citenamefont {Fiore}, \citenamefont
  {Cingolani}, \citenamefont {Manna}, \citenamefont {Cola},\ and\ \citenamefont
  {Krahne}}]{Persano2010}%
  \BibitemOpen
  \bibfield  {author} {\bibinfo {author} {\bibfnamefont {A.}~\bibnamefont
  {Persano}}, \bibinfo {author} {\bibfnamefont {M.~D.}\ \bibnamefont {Giorgi}},
  \bibinfo {author} {\bibfnamefont {A.}~\bibnamefont {Fiore}}, \bibinfo
  {author} {\bibfnamefont {R.}~\bibnamefont {Cingolani}}, \bibinfo {author}
  {\bibfnamefont {L.}~\bibnamefont {Manna}}, \bibinfo {author} {\bibfnamefont
  {A.}~\bibnamefont {Cola}},\ and\ \bibinfo {author} {\bibfnamefont
  {R.}~\bibnamefont {Krahne}},\ }\bibfield  {title} {\bibinfo {title}
  {Photoconduction properties in aligned assemblies of colloidal {CdSe}/{CdS}
  nanorods},\ }\href {https://doi.org/10.1021/nn901575r} {\bibfield  {journal}
  {\bibinfo  {journal} {{ACS} Nano}\ }\textbf {\bibinfo {volume} {4}},\
  \bibinfo {pages} {1646} (\bibinfo {year} {2010})}\BibitemShut {NoStop}%
\bibitem [{\citenamefont {Aslan}\ \emph {et~al.}(2005)\citenamefont {Aslan},
  \citenamefont {Leonenko}, \citenamefont {Lakowicz},\ and\ \citenamefont
  {Geddes}}]{Aslan2005}%
  \BibitemOpen
  \bibfield  {author} {\bibinfo {author} {\bibfnamefont {K.}~\bibnamefont
  {Aslan}}, \bibinfo {author} {\bibfnamefont {Z.}~\bibnamefont {Leonenko}},
  \bibinfo {author} {\bibfnamefont {J.~R.}\ \bibnamefont {Lakowicz}},\ and\
  \bibinfo {author} {\bibfnamefont {C.~D.}\ \bibnamefont {Geddes}},\ }\bibfield
   {title} {\bibinfo {title} {Fast and slow deposition of silver nanorods on
  planar surfaces: {Application} to metal-enhanced fluorescence},\ }\href
  {https://doi.org/10.1021/jp045186t} {\bibfield  {journal} {\bibinfo
  {journal} {J. Phys. Chem. B}\ }\textbf {\bibinfo {volume} {109}},\ \bibinfo
  {pages} {3157} (\bibinfo {year} {2005})}\BibitemShut {NoStop}%
\bibitem [{\citenamefont {Talbot}\ \emph {et~al.}(2000)\citenamefont {Talbot},
  \citenamefont {{Van Tassel}},\ and\ \citenamefont {Viot}}]{Talbot2000}%
  \BibitemOpen
  \bibfield  {author} {\bibinfo {author} {\bibfnamefont {G.}~\bibnamefont
  {Talbot}, \bibfnamefont {J.~and.~Tarjus}}, \bibinfo {author} {\bibfnamefont
  {P.~R.}\ \bibnamefont {{Van Tassel}}},\ and\ \bibinfo {author} {\bibfnamefont
  {P.}~\bibnamefont {Viot}},\ }\bibfield  {title} {\bibinfo {title} {From car
  parking to protein adsorption: {An} overview of sequential adsorption
  processes},\ }\href {https://doi.org/10.1016/S0927-7757(99)00409-4}
  {\bibfield  {journal} {\bibinfo  {journal} {Colloids Surf. A: Physicochem.
  Eng. Asp.}\ }\textbf {\bibinfo {volume} {165}},\ \bibinfo {pages} {287}
  (\bibinfo {year} {2000})}\BibitemShut {NoStop}%
\bibitem [{\citenamefont {Onsager}(1949)}]{Onsager1949ANYAS}%
  \BibitemOpen
  \bibfield  {author} {\bibinfo {author} {\bibfnamefont {L.}~\bibnamefont
  {Onsager}},\ }\bibfield  {title} {\bibinfo {title} {The effects of shape on
  the interaction of colloidal particles},\ }\href
  {https://doi.org/10.1111/j.1749-6632.1949.tb27296.x} {\bibfield  {journal}
  {\bibinfo  {journal} {Ann. N. Y. Acad. Sci.}\ }\textbf {\bibinfo {volume}
  {51}},\ \bibinfo {pages} {627} (\bibinfo {year} {1949})}\BibitemShut
  {NoStop}%
\bibitem [{\citenamefont {Leroyer}\ and\ \citenamefont
  {Pommiers}(1994)}]{Leroyer1994PRB}%
  \BibitemOpen
  \bibfield  {author} {\bibinfo {author} {\bibfnamefont {Y.}~\bibnamefont
  {Leroyer}}\ and\ \bibinfo {author} {\bibfnamefont {E.}~\bibnamefont
  {Pommiers}},\ }\bibfield  {title} {\bibinfo {title} {Monte {Carlo} analysis
  of percolation of line segments on a square lattice},\ }\href
  {https://doi.org/10.1103/PhysRevB.50.2795} {\bibfield  {journal} {\bibinfo
  {journal} {Phys. Rev. B}\ }\textbf {\bibinfo {volume} {50}},\ \bibinfo
  {pages} {2795} (\bibinfo {year} {1994})}\BibitemShut {NoStop}%
\bibitem [{\citenamefont {Kondrat}\ and\ \citenamefont
  {P\c{e}kalski}(2001)}]{Kondrat2001PRE}%
  \BibitemOpen
  \bibfield  {author} {\bibinfo {author} {\bibfnamefont {G.}~\bibnamefont
  {Kondrat}}\ and\ \bibinfo {author} {\bibfnamefont {A.}~\bibnamefont
  {P\c{e}kalski}},\ }\bibfield  {title} {\bibinfo {title} {Percolation and
  jamming in random sequential adsorption of linear segments on a square
  lattice},\ }\href {https://doi.org/10.1103/PhysRevE.63.051108} {\bibfield
  {journal} {\bibinfo  {journal} {Phys. Rev. E}\ }\textbf {\bibinfo {volume}
  {63}},\ \bibinfo {pages} {051108} (\bibinfo {year} {2001})}\BibitemShut
  {NoStop}%
\bibitem [{\citenamefont {Adamczyk}\ \emph {et~al.}(2008)\citenamefont
  {Adamczyk}, \citenamefont {Romiszowski},\ and\ \citenamefont
  {Sikorski}}]{Adamczyk2008JChemPhys}%
  \BibitemOpen
  \bibfield  {author} {\bibinfo {author} {\bibfnamefont {P.}~\bibnamefont
  {Adamczyk}}, \bibinfo {author} {\bibfnamefont {P.}~\bibnamefont
  {Romiszowski}},\ and\ \bibinfo {author} {\bibfnamefont {A.}~\bibnamefont
  {Sikorski}},\ }\bibfield  {title} {\bibinfo {title} {A simple model of stiff
  and flexible polymer chain adsorption: {The} influence of the internal chain
  architecture},\ }\href {https://doi.org/10.1063/1.2907715} {\bibfield
  {journal} {\bibinfo  {journal} {J. Chem. Phys.}\ }\textbf {\bibinfo {volume}
  {128}},\ \bibinfo {pages} {154911} (\bibinfo {year} {2008})}\BibitemShut
  {NoStop}%
\bibitem [{\citenamefont {Manna}\ and\ \citenamefont
  {\v{S}vraki{\'c}}(1991)}]{Manna1991JPhA}%
  \BibitemOpen
  \bibfield  {author} {\bibinfo {author} {\bibfnamefont {S.~S.}\ \bibnamefont
  {Manna}}\ and\ \bibinfo {author} {\bibfnamefont {N.~M.}\ \bibnamefont
  {\v{S}vraki{\'c}}},\ }\bibfield  {title} {\bibinfo {title} {Random sequential
  adsorption: line segments on the square lattice},\ }\href
  {https://doi.org/10.1088/0305-4470/24/12/003} {\bibfield  {journal} {\bibinfo
   {journal} {J. Phys. A: Math. Gen.}\ }\textbf {\bibinfo {volume} {24}},\
  \bibinfo {pages} {L671} (\bibinfo {year} {1991})}\BibitemShut {NoStop}%
\bibitem [{\citenamefont {Budinski-Petkovi\'{c}}\ and\ \citenamefont
  {Kozmidis-Luburi\'{c}}(1997{\natexlab{a}})}]{BudinskiPetkovic1997PhysA}%
  \BibitemOpen
  \bibfield  {author} {\bibinfo {author} {\bibfnamefont {L.}~\bibnamefont
  {Budinski-Petkovi\'{c}}}\ and\ \bibinfo {author} {\bibfnamefont
  {U.}~\bibnamefont {Kozmidis-Luburi\'{c}}},\ }\bibfield  {title} {\bibinfo
  {title} {Jamming configurations for irreversible deposition on a square
  lattice},\ }\href {https://doi.org/10.1016/S0378-4371(96)00374-3} {\bibfield
  {journal} {\bibinfo  {journal} {Physica A}\ }\textbf {\bibinfo {volume}
  {236}},\ \bibinfo {pages} {211} (\bibinfo {year}
  {1997}{\natexlab{a}})}\BibitemShut {NoStop}%
\bibitem [{\citenamefont {Vandewalle}\ \emph {et~al.}(2000)\citenamefont
  {Vandewalle}, \citenamefont {Galam},\ and\ \citenamefont
  {Kramer}}]{Vandewalle2000EPJE}%
  \BibitemOpen
  \bibfield  {author} {\bibinfo {author} {\bibfnamefont {N.}~\bibnamefont
  {Vandewalle}}, \bibinfo {author} {\bibfnamefont {S.}~\bibnamefont {Galam}},\
  and\ \bibinfo {author} {\bibfnamefont {M.}~\bibnamefont {Kramer}},\
  }\bibfield  {title} {\bibinfo {title} {A new universality for random
  sequential deposition of needles},\ }\href
  {https://doi.org/10.1007/s100510051047} {\bibfield  {journal} {\bibinfo
  {journal} {Eur. Phys. J. B}\ }\textbf {\bibinfo {volume} {14}},\ \bibinfo
  {pages} {407} (\bibinfo {year} {2000})}\BibitemShut {NoStop}%
\bibitem [{\citenamefont {Budinski-Petkovi\'{c}}\ and\ \citenamefont
  {Kozmidis-Luburi\'{c}}(1997{\natexlab{b}})}]{BudinskiPetkovic1997PRE}%
  \BibitemOpen
  \bibfield  {author} {\bibinfo {author} {\bibfnamefont {L.}~\bibnamefont
  {Budinski-Petkovi\'{c}}}\ and\ \bibinfo {author} {\bibfnamefont
  {U.}~\bibnamefont {Kozmidis-Luburi\'{c}}},\ }\bibfield  {title} {\bibinfo
  {title} {Random sequential adsorption on a triangular lattice},\ }\href
  {https://doi.org/10.1103/PhysRevE.56.6904} {\bibfield  {journal} {\bibinfo
  {journal} {Phys. Rev. E}\ }\textbf {\bibinfo {volume} {56}},\ \bibinfo
  {pages} {6904} (\bibinfo {year} {1997}{\natexlab{b}})}\BibitemShut {NoStop}%
\bibitem [{\citenamefont {Lee}(2000)}]{Lee2000}%
  \BibitemOpen
  \bibfield  {author} {\bibinfo {author} {\bibfnamefont {J.~W.}\ \bibnamefont
  {Lee}},\ }\bibfield  {title} {\bibinfo {title} {Irreversible random
  sequential adsorption of mixtures},\ }\href
  {https://doi.org/10.1016/S0927-7757(99)00414-8} {\bibfield  {journal}
  {\bibinfo  {journal} {Colloids Surf. A: Physicochem. Eng. Asp.}\ }\textbf
  {\bibinfo {volume} {165}},\ \bibinfo {pages} {363} (\bibinfo {year}
  {2000})}\BibitemShut {NoStop}%
\bibitem [{\citenamefont {Tarasevich}\ \emph {et~al.}(2019)\citenamefont
  {Tarasevich}, \citenamefont {Eserkepov}, \citenamefont {Chirkova},\ and\
  \citenamefont {Goltseva}}]{Tarasevich2019JPCS}%
  \BibitemOpen
  \bibfield  {author} {\bibinfo {author} {\bibfnamefont {Y.~Y.}\ \bibnamefont
  {Tarasevich}}, \bibinfo {author} {\bibfnamefont {A.~V.}\ \bibnamefont
  {Eserkepov}}, \bibinfo {author} {\bibfnamefont {V.~V.}\ \bibnamefont
  {Chirkova}},\ and\ \bibinfo {author} {\bibfnamefont {V.~A.}\ \bibnamefont
  {Goltseva}},\ }\bibfield  {title} {\bibinfo {title} {{Monte} {Carlo}
  simulation of entropy-driven pattern formation in a two-dimensional system of
  rectangular particles},\ }\href
  {https://doi.org/10.1088/1742-6596/1163/1/012007} {\bibfield  {journal}
  {\bibinfo  {journal} {J. Phys. Conf. Ser.}\ }\textbf {\bibinfo {volume}
  {1163}},\ \bibinfo {pages} {012007} (\bibinfo {year} {2019})}\BibitemShut
  {NoStop}%
\bibitem [{\citenamefont {Slutskii}\ \emph {et~al.}(2018)\citenamefont
  {Slutskii}, \citenamefont {Barash},\ and\ \citenamefont
  {Tarasevich}}]{Slutskii2018PRE}%
  \BibitemOpen
  \bibfield  {author} {\bibinfo {author} {\bibfnamefont {M.~G.}\ \bibnamefont
  {Slutskii}}, \bibinfo {author} {\bibfnamefont {L.~Y.}\ \bibnamefont
  {Barash}},\ and\ \bibinfo {author} {\bibfnamefont {Y.~Y.}\ \bibnamefont
  {Tarasevich}},\ }\bibfield  {title} {\bibinfo {title} {Percolation and
  jamming of random sequential adsorption samples of large linear $k$-mers on a
  square lattice},\ }\href {https://doi.org/10.1103/PhysRevE.98.062130}
  {\bibfield  {journal} {\bibinfo  {journal} {Phys. Rev. E}\ }\textbf {\bibinfo
  {volume} {98}},\ \bibinfo {pages} {062130} (\bibinfo {year}
  {2018})}\BibitemShut {NoStop}%
\bibitem [{\citenamefont {Ricci}\ \emph {et~al.}(1994)\citenamefont {Ricci},
  \citenamefont {Talbot}, \citenamefont {Tarjus},\ and\ \citenamefont
  {Viot}}]{Ricci1994JChemPhys}%
  \BibitemOpen
  \bibfield  {author} {\bibinfo {author} {\bibfnamefont {S.~M.}\ \bibnamefont
  {Ricci}}, \bibinfo {author} {\bibfnamefont {J.}~\bibnamefont {Talbot}},
  \bibinfo {author} {\bibfnamefont {G.}~\bibnamefont {Tarjus}},\ and\ \bibinfo
  {author} {\bibfnamefont {P.}~\bibnamefont {Viot}},\ }\bibfield  {title}
  {\bibinfo {title} {A structural comparison of random sequential adsorption
  and equilibrium configurations of spherocylinders},\ }\href
  {https://doi.org/10.1063/1.468046} {\bibfield  {journal} {\bibinfo  {journal}
  {J. Chem. Phys.}\ }\textbf {\bibinfo {volume} {101}},\ \bibinfo {pages}
  {9164} (\bibinfo {year} {1994})}\BibitemShut {NoStop}%
\bibitem [{\citenamefont {Hart}\ and\ \citenamefont
  {Aar{\~a}o~Reis}(2016)}]{Hart2016PRE}%
  \BibitemOpen
  \bibfield  {author} {\bibinfo {author} {\bibfnamefont {R.~C.}\ \bibnamefont
  {Hart}}\ and\ \bibinfo {author} {\bibfnamefont {F.~D.~A.}\ \bibnamefont
  {Aar{\~a}o~Reis}},\ }\bibfield  {title} {\bibinfo {title} {Random sequential
  adsorption of polydisperse mixtures on lattices},\ }\href
  {https://doi.org/10.1103/PhysRevE.94.022802} {\bibfield  {journal} {\bibinfo
  {journal} {Phys. Rev. E}\ }\textbf {\bibinfo {volume} {94}},\ \bibinfo
  {pages} {022802} (\bibinfo {year} {2016})}\BibitemShut {NoStop}%
\bibitem [{\citenamefont {Kundu}\ \emph {et~al.}(2013)\citenamefont {Kundu},
  \citenamefont {Rajesh}, \citenamefont {Dhar},\ and\ \citenamefont
  {Stilck}}]{Kundu2013PRE}%
  \BibitemOpen
  \bibfield  {author} {\bibinfo {author} {\bibfnamefont {J.}~\bibnamefont
  {Kundu}}, \bibinfo {author} {\bibfnamefont {R.}~\bibnamefont {Rajesh}},
  \bibinfo {author} {\bibfnamefont {D.}~\bibnamefont {Dhar}},\ and\ \bibinfo
  {author} {\bibfnamefont {J.~F.}\ \bibnamefont {Stilck}},\ }\bibfield  {title}
  {\bibinfo {title} {Nematic-disordered phase transition in systems of long
  rigid rods on two-dimensional lattices},\ }\href
  {https://doi.org/10.1103/PhysRevE.87.032103} {\bibfield  {journal} {\bibinfo
  {journal} {Phys. Rev. E}\ }\textbf {\bibinfo {volume} {87}},\ \bibinfo
  {pages} {032103} (\bibinfo {year} {2013})}\BibitemShut {NoStop}%
\bibitem [{\citenamefont {Tarasevich}\ \emph {et~al.}(2017)\citenamefont
  {Tarasevich}, \citenamefont {Laptev}, \citenamefont {Burmistrov},\ and\
  \citenamefont {Lebovka}}]{Tarasevich2017JSM}%
  \BibitemOpen
  \bibfield  {author} {\bibinfo {author} {\bibfnamefont {Y.~Y.}\ \bibnamefont
  {Tarasevich}}, \bibinfo {author} {\bibfnamefont {V.~V.}\ \bibnamefont
  {Laptev}}, \bibinfo {author} {\bibfnamefont {A.~S.}\ \bibnamefont
  {Burmistrov}},\ and\ \bibinfo {author} {\bibfnamefont {N.~I.}\ \bibnamefont
  {Lebovka}},\ }\bibfield  {title} {\bibinfo {title} {Pattern formation in a
  two-dimensional two-species diffusion model with anisotropic nonlinear
  diffusivities: a lattice approach},\ }\href
  {https://doi.org/10.1088/1742-5468/aa82bf} {\bibfield  {journal} {\bibinfo
  {journal} {J. Stat. Mech: Theory Exp.}\ }\textbf {\bibinfo {volume} {2017}},\
  \bibinfo {pages} {093203} (\bibinfo {year} {2017})}\BibitemShut {NoStop}%
\bibitem [{\citenamefont {Sherwood}(1990)}]{Sherwood1990JPhA}%
  \BibitemOpen
  \bibfield  {author} {\bibinfo {author} {\bibfnamefont {J.~D.}\ \bibnamefont
  {Sherwood}},\ }\bibfield  {title} {\bibinfo {title} {Random sequential
  adsorption of lines and ellipses},\ }\href
  {https://doi.org/10.1088/0305-4470/23/13/021} {\bibfield  {journal} {\bibinfo
   {journal} {J. Phys. A: Math. Gen.}\ }\textbf {\bibinfo {volume} {23}},\
  \bibinfo {pages} {2827} (\bibinfo {year} {1990})}\BibitemShut {NoStop}%
\bibitem [{\citenamefont {Vigil}\ and\ \citenamefont
  {Ziff}(1989)}]{Vigil1989JChemPhys}%
  \BibitemOpen
  \bibfield  {author} {\bibinfo {author} {\bibfnamefont {R.~D.}\ \bibnamefont
  {Vigil}}\ and\ \bibinfo {author} {\bibfnamefont {R.~M.}\ \bibnamefont
  {Ziff}},\ }\bibfield  {title} {\bibinfo {title} {Random sequential adsorption
  of unoriented rectangles onto a plane},\ }\href
  {https://doi.org/10.1063/1.457021} {\bibfield  {journal} {\bibinfo  {journal}
  {J. Chem. Phys.}\ }\textbf {\bibinfo {volume} {91}},\ \bibinfo {pages} {2599}
  (\bibinfo {year} {1989})}\BibitemShut {NoStop}%
\bibitem [{\citenamefont {Cie\'{s}la}(2013)}]{Ciesla2013PRE}%
  \BibitemOpen
  \bibfield  {author} {\bibinfo {author} {\bibfnamefont {M.}~\bibnamefont
  {Cie\'{s}la}},\ }\bibfield  {title} {\bibinfo {title} {Continuum random
  sequential adsorption of polymer on a flat and homogeneous surface},\ }\href
  {https://doi.org/10.1103/PhysRevE.87.052401} {\bibfield  {journal} {\bibinfo
  {journal} {Phys. Rev. E}\ }\textbf {\bibinfo {volume} {87}},\ \bibinfo
  {pages} {052401} (\bibinfo {year} {2013})}\BibitemShut {NoStop}%
\bibitem [{\citenamefont {Tsiantis}\ and\ \citenamefont
  {Papathanasiou}(2019)}]{Tsiantis2019PhysA}%
  \BibitemOpen
  \bibfield  {author} {\bibinfo {author} {\bibfnamefont {A.}~\bibnamefont
  {Tsiantis}}\ and\ \bibinfo {author} {\bibfnamefont {T.~D.}\ \bibnamefont
  {Papathanasiou}},\ }\bibfield  {title} {\bibinfo {title} {A novel {FastRSA}
  algorithm: {Statistical} properties and evolution of microstructure},\ }\href
  {https://doi.org/10.1016/j.physa.2019.122083} {\bibfield  {journal} {\bibinfo
   {journal} {Physica A}\ }\textbf {\bibinfo {volume} {534}},\ \bibinfo {pages}
  {122083} (\bibinfo {year} {2019})}\BibitemShut {NoStop}%
\bibitem [{\citenamefont {Heitz}\ \emph {et~al.}(2011)\citenamefont {Heitz},
  \citenamefont {Leroy}, \citenamefont {H{\'{e}}brard},\ and\ \citenamefont
  {Lallement}}]{Heitz2011NanoTech}%
  \BibitemOpen
  \bibfield  {author} {\bibinfo {author} {\bibfnamefont {J.}~\bibnamefont
  {Heitz}}, \bibinfo {author} {\bibfnamefont {Y.}~\bibnamefont {Leroy}},
  \bibinfo {author} {\bibfnamefont {L.}~\bibnamefont {H{\'{e}}brard}},\ and\
  \bibinfo {author} {\bibfnamefont {C.}~\bibnamefont {Lallement}},\ }\bibfield
  {title} {\bibinfo {title} {Theoretical characterization of the topology of
  connected carbon nanotubes in random networks},\ }\href
  {https://doi.org/10.1088/0957-4484/22/34/345703} {\bibfield  {journal}
  {\bibinfo  {journal} {Nanotechnology}\ }\textbf {\bibinfo {volume} {22}},\
  \bibinfo {pages} {345703} (\bibinfo {year} {2011})}\BibitemShut {NoStop}%
\bibitem [{Note1()}]{Note1}%
  \BibitemOpen
  \bibinfo {note} {See Supplemental Material at [URL will be inserted by
  publisher] for an animation of the temporal evolution of domains for $k=8$,
  $L=16k$.}\BibitemShut {Stop}%
\end{thebibliography}%

\end{document}